\documentclass[a4paper]{article}
\usepackage{cite}
\usepackage[dvips]{graphicx}
\usepackage{hhline}

\begin{document}

\title{\center{P- and T-violating Schiff moment of the Mercury nucleus}}
\author{V.F. Dmitriev and R.A. Sen'kov \\
{\it Budker Institute of Nuclear Physics }\\
{\it and }\\
{\it Novosibirsk State University}}
\maketitle
\begin{abstract}
The Schiff moment of the $^{199}$Hg nucleus was calculated using finite range
P- and T-violating weak nucleon-nucleon interaction. Effects of the core
polarization were considered in the framework of RPA with effective residual
forces.
\end{abstract}
\section{Introduction}
The most precise limit on parity- and time-invariance violating
nucleon-nucleon interaction has been obtained from the
measurement of the atomic electric dipole moment of $^{199}$Hg
\cite{rgf01}. The hadronic part of the atomic dipole moment associated with the
electric dipole moment of the $^{199}$Hg nucleus manifests itself through
the Schiff moment which is the first nonzero term in the
expansion of the nuclear electromagnetic potential after
including the screening of the atomic electrons
\cite{pr50,sch63,san67}.

The operator for the Schiff moment is \cite{fks84}
\begin{equation} \label{1}
S_\mu = \frac{1}{10}\sqrt{\frac{4\pi}{3}}\sum_i^A e_i \left(r_i^3 -
\frac{5}{3}\langle r^2\rangle_{ch} r_i\right) Y_{1\mu}(\hat{{\bf r}}_i),
\end{equation}
 where $e_i$ is $e$ for a proton and zero for a neutron. The Schiff moment
generates T- and P-odd electrostatic potential in the form
\begin{equation} \label{2}
\phi({\bf r}) = 4\pi {\bf S}\cdot\mbox{\boldmath $\nabla$}\delta({\bf r}).
\end{equation}
Interaction of atomic electrons with the potential given by Eq. (\ref{2})
produces an atomic dipole moment
 \begin{equation} \label{3}
d_{atom} = \sum_n\frac{\langle 0| -e\sum_i^Z \phi({\bf
r}_i)|n\rangle\langle n|-e\sum_i^Z z_i|0\rangle}{E_n-E_0} + {\rm h.c.}
\end{equation}
Due to contact origin of the potential, the electrons in $s$- and
$p$- atomic orbitals only contribute to the dipole moment given by
Eq. (\ref{3}).

The nuclear Schiff moment was calculated so far in a simplified
model \cite{fks84,fg02} without considering many-body nuclear
structure effects. These effects have to be understood properly
if we intend to extract the parameters of P- and T-violating
nuclear interaction from the value of the Schiff moment.  For
light nuclei the properties of the Schiff moment strength
obtained in modern shell model calculations were discussed in
Ref. \cite{efh00}. The study of the polarization effects
associated with the coupling to isoscalar dipole compression mode
has been performed in Ref. \cite{hb00}.  In our paper we
calculate the Schiff moment of $^{199}$Hg nucleus within RPA
framework with effective residual strong forces using finite
range P- and T-odd weak nuclear interaction.
\section{Basic ingredients of the theory}
\subsection{Nucleon-nucleon P- and T-odd interaction}
We use the interaction generated by P- and T-violating pion exchange
\cite{hh83,h88,khk00}.
$$
W({\bf r}_1 - {\bf r}_2) = -\frac{g}{8\pi m_p}\left[ (g_0\mbox{\boldmath
 $\tau$}_1\cdot\mbox{\boldmath $\tau$}_2   +
g_2(\mbox{\boldmath $\tau$}_1\cdot \mbox{\boldmath $\tau$}_2
 -3\tau_1^3\tau_2^3))   ( \mbox{\boldmath $\sigma$}_1 -
\mbox{\boldmath $\sigma$}_2)\right.
$$
\begin{equation} \label{4}
\left. + g_1(\tau_1^3\mbox{\boldmath $\sigma$}_1  - \tau_2^3\mbox{\boldmath
$\sigma$}_2)\right]  \cdot \mbox{\boldmath $\nabla$}_1 \frac{e^{-m_{\pi}
r_{12}}}{r_{12}},
\end{equation}
where $g$ is the usual strong pion-nucleon pseudoscalar coupling
constant, $g_0$, $g_1$, and $g_2$ correspond to  isoscalar, isovector , and
isotensor P- and T-odd couplings, $m_p$ is the proton mass.
In contrast to P-odd and T-even interaction, in Eq. (\ref{4}) the  exchange of
$\pi^0$ is allowed. This term produces the direct contribution to P- and T-
odd part of nuclear mean field while the other terms produce the
exchange contribution only. Since the direct contribution
dominates for finite range potentials we can expect that the
interaction (\ref{4}) is the leading one and the exchange of
heavier mesons can be omitted.

In previous calculations the phenomenological contact interaction
has often been used instead of finite range interaction given by
Eq. (\ref{4}). It has the form \cite{kh91}
$$
W_c( {\bf r}_a - {\bf r}_b) = \frac{G}{\sqrt{2}} \frac{1}{2m_p} \left(
(\eta_{ab}\mbox{\boldmath $\sigma$}_a - \eta_{ba} \mbox{\boldmath
$\sigma$}_b)\cdot \mbox{ \boldmath $\nabla$}_a \delta ({\bf r}_a - {\bf r}_b) +
\right.
$$
\begin{equation} \label{5}
\left. \eta_{ab}' [ \mbox{\boldmath $\sigma$}_a \times \mbox{\boldmath
$\sigma$}_b]\cdot\{({\bf p}_a - {\bf p}_b),\delta ({\bf r}_a - {\bf
r}_b)\}\right),
\end{equation}
where $G$ is the Fermi constant. In the limit $m_\pi \rightarrow
\infty$  the interaction (\ref{4}) transforms into Eq. (\ref{5})
after the substitution $gg_i \rightarrow
\frac{Gm_\pi^2}{\sqrt{2}}\eta$. We shall use this factor when
comparing our results with those obtained using the contact
interaction given by Eq. (\ref{5}).

\subsection{Nuclear mean field and correction from the weak forces}
In our calculations we used full single-particle spectrum including continuum.
The single-particle basis was obtained using partially self-consistent
mean-field potential of Ref. \cite{bs74}. The potential includes four terms. The
isoscalar term is the standard Woods-Saxon potential
\begin{equation} \label{6}
U_0(r) = - \frac{V}{1+\exp{\frac{r-R}{a}}},
\end{equation}
with the parameters $V = 52.03$ MeV, $R = 1.2709 A^{1/3}$ fm, and
$a = 0.742$ fm. Two other terms  $U_{ls}(r)$, and $U_\tau(r)$ were calculated
self-consistently using two-body Migdal-type interaction of Ref. \cite{m67} for
the spin-orbit and isovector parts of the potential. The last term is the
Coulomb potential calculated for uniformly charged sphere  with $R_C =
1.18 A^{1/3}$ fm. The mean field potential obtained in this way produces good
fit for single particle energies and r.m.s. radii for nuclei in the lead region.

The correction to the mean field (\ref{6}) from the weak
interaction (\ref{4}) consist of direct and exchange terms. The
direct term has the form
$$
 \delta U_{dir}({\bf r}) = \frac{gm_\pi^2}{\pi m_p} (\mbox{\boldmath
$\sigma$}\cdot {\bf n})\tau^3 \int_0^\infty r'^2dr'\;
b_{1\,0}(r,r')[(g_0-2g_2)(\rho_p(r)-\rho_n(r)) +
$$
\begin{equation} \label{7}
 g_1( \rho_p(r)+\rho_n(r))],
\end{equation}
where the function $b_{1\,0}(r_1,r_2)$ is a combination of spherical Bessel
functions of imaginary argument
\begin{equation} \label{8}
b_{l_1l_2} (r_1,r_2) = i_{l_1}(m_\pi r_1)k_{l_2}(m_\pi r_2) \theta(r_2-r_1) -
i_{l_2}(m_\pi r_2)k_{l_1}(m_\pi r_1) \theta(r_1-r_2).
\end{equation}
Note, that the potential given by Eq. (\ref{7}) is pure isovector.
The contribution of the isovector interaction component dominates in Eq.
(\ref{7}),  the isoscalar and isotensor components of the interaction are
suppressed by the factor $\frac{N-Z}{A}$.
For zero range interaction the potential would be proportional to ${\mathbf
\nabla} \rho(r)$. The gradient makes this potential very sensitive to the
details of nuclear surface. Our potential given by Eq. (\ref{7})
is less sensitive to the surface due  to additional integration
over region of the order of pion Compton wavelength.

The exchange term is more complicated. The matrix element of it
taken over angular variable is a non-local operator in radial
coordinates.
$$
\langle\tilde{\nu}|\delta U_{exch}({\bf r},{\bf r}')|\nu\rangle=W_\nu(r,r'),
$$
where $|\tilde{\nu}\rangle = -(\mbox{\boldmath $\sigma$}\cdot {\bf
n})|\nu\rangle$, and $$
 W_\nu(r,r') = \frac{1}{2j_\nu+1}\frac{gm_\pi^2}{\pi m_p}Tr_2\left\{
\sum_{\kappa l_1 l_2} \left(\frac{g_0}{2}(3-\tau_1^3\tau_2^3)-2g_2
\tau_1^3\tau_2^3 + \frac{g_1}{2}(\tau_1^3 +\tau_2^3)\right) \times \right.
$$
$$
n_\kappa \left(\begin{array}{ccc}
l_1 & l_2 & 1 \\ 0 & 0 & 0 \end{array}\right) b_{l_1 l_2}(r,r')R_\kappa(r)R_\kappa(r')
\left[
(-)^{l_1}[l_1](\kappa||T_{l_2}^{l_1}||\tilde{\nu})^*(\kappa||Y_{l_2}||\nu)
-\right.
$$
 \begin{equation} \label{9}
\left.\left.
(-)^{l_2}[l_2](\tilde{\nu}||Y_{l_1}||\kappa)(\nu||T_{l_1}^{l_2}||\kappa)^*\right]\right\}.
\end{equation}
The Trace is assumed over isospin variable of the second particle
, $n_\kappa$ is the occupation number of the single particle
state $|\kappa\rangle$, $[l]=\sqrt{2l+1}$ , and the tensor operator
$T^L_{JM}({\bf n}) = \{\mbox{\boldmath $\sigma$}\otimes Y_L ({\bf
n})\}_{JM}$.

The correction to the single particle wave function $\psi_\nu({\bf r})$ can be
presented as
$$
\delta \psi_\nu({\bf r}) = (\mbox{\boldmath $\sigma$}\cdot {\bf n})
\Omega_\nu({\bf n}) \delta R_\nu(r),
$$
where $\Omega_\nu({\bf n})$ is the angular part of the wave
function. The radial correction $\delta R_\nu (r)$ is the sum of
the direct and the exchange terms $\delta R_\nu (r)=\delta
R_{\nu\,dir}(r) +\delta R_{\nu\,exch}(r)$, where
$$
 \delta R_{\nu\,dir}(r) = \int_0^\infty {\mbox{\sl
G}}_{j_\nu \tilde{l}_\nu}(r,r'|\epsilon_\nu)\delta
U_{dir}(r')R_\nu(r') r'^2\,dr' , $$
\begin{equation} \label{10}
 \delta R_{\nu\,exch}(r) = \int_0^\infty  {\mbox{\sl
G}}_{j_\nu \tilde{l}_\nu}(r,r'|\epsilon_\nu)
W_\nu(r',r'')R_\nu(r'')r'^2r''^2\,dr'dr''.
\end{equation}
Here $  {\mbox{\sl G}}_{j_\nu \tilde{l}_\nu}(r,r'|\epsilon_\nu)  $
is the Green function of radial Schr\"odinger equation for the
total angular momentum $j_\nu$, and  the  orbital angular momentum
$\tilde{l}_\nu = 2j_\nu - l_\nu$ . $ \epsilon_\nu  $ is the single
particle energy.

\section{Core polarization}
The effects of the core polarization for a one particle operator can be treated
introducing a renormalized operator $\tilde{{\bf S}}$ satisfying the equation
 \begin{equation} \label{11}
 \tilde{\bf S}_{\nu'\nu}  = {\bf S}^0_{\nu'\nu} + \sum_{\mu'\mu}  \tilde{\bf
S}_{\mu\mu'} \frac{n_\mu -n_{\mu'}}{\epsilon_\mu - \epsilon_{\mu'}+\omega}
\langle \nu'\mu'|F+W|\mu\nu\rangle,
\end{equation}
where ${\bf S}^0$ is the bare Schiff moment operator given by Eq.
(\ref{1}), $n_\mu$  and $\epsilon_\mu$ are the single particle
occupation numbers and energies.  For static moments the external frequency
$\omega \rightarrow 0$.   The interaction  in Eq.
(\ref{11}) includes both the strong residual interaction $F$ and the
weak one. The latter we take in the form given by Eq. (\ref{4}).
Strictly speaking, the interaction of two nucleons in nuclear
matter differs from the interaction in the vacuum. We do not
discuss this effect here and keep the weak interaction in the form
(\ref{4}).  The single particle wave functions in Eq. (\ref{11})
are the eigenstates of the mean field which is also the sum of
the strong and the weak fields given by Eqs. (\ref{7},\ref{9}).
Since the weak forces are really small compared to the strong
interaction, it is natural to treat them perturbatively. The
simplest way to do it is to present $<{\bf r}|\nu> = \psi_\nu
({\bf r}) + \delta\psi_\nu ({\bf r})$ and to gather the terms
linear in $\delta\psi$ and $W({\bf r}_1-{\bf r}_2)$. It is
convenient to present the matrix element of the Schiff moment as
a sum of three terms
\begin{equation} \label{12}
\tilde{\bf S}_{\nu\nu} = \langle \delta\psi_\nu|{\bf S}|\psi_\nu\rangle +
\langle \psi_\nu|{\bf S}|\delta\psi_\nu\rangle + \langle \psi_\nu|\delta{\bf
S}|\psi_\nu\rangle.
\end{equation}
The operator $\bf S$ in the first two terms satisfies the same
Eq. (\ref{11}) where the strong interaction only is included. The
corrections from the weak forces for these terms are in the wave
functions of an odd nucleon only.   The third term presents an
induced contribution \cite{dt97} arising from the P- and
T-violating corrections to the intermediate states $|\mu\rangle$
and $|\mu'\rangle$. The equation for $\delta {\bf S}$ is
\begin{equation} \label{13}
(\delta{\bf S}-\delta{\bf S}_{NL})_{\nu'\nu} =  (\delta{\bf S}_0)_{\nu'\nu} +
\sum_{\mu'\mu} (\delta{\bf S}-\delta{\bf S}_{NL})_{\mu\mu'} \frac{n_\mu -n_{\mu'}}{\epsilon_\mu -
\epsilon_{\mu'}} \langle \nu'\mu'|F|\mu\nu\rangle,
\end{equation}
where  $\delta{\bf S}_{NL}$  is the non-local part of
$\delta{\bf S}$ produced by the exchange matrix elements of weak
interaction $W$.
 \begin{equation}\label{AS3}
({\bf \delta S_{NL}})_{\nu'\nu} = \sum_{\mu
\mu'}
({\bf S})_{\mu \mu'}\frac{n_\mu -n_{\mu'}}{\epsilon_\mu-\epsilon_{\mu'}}
\left< \nu' \mu'| W_{exch}|\mu\nu \right>.
\end{equation}
The equation for $\delta{\bf S}_0$ is
$$
(\delta{\bf S}_0)_{\nu'\nu} = \sum_{\mu\mu'} \frac{n_\mu -n_{\mu'}}{\epsilon_\mu
- \epsilon_{\mu'}}\left( \langle \mu|{\bf
S}|\mu'\rangle\langle\nu'\mu'|W_{dir}|\mu\nu\rangle + \langle \mu|\delta {\bf
S}_{NL}|\mu'\rangle\langle\nu'\mu'|F|\mu\nu\rangle + \right.
$$
$$
\langle \delta\psi_\mu|{\bf
S}|\mu'\rangle\langle\nu'\mu'|F|\mu\nu\rangle + \langle \mu|{\bf
S}|\delta\psi_{\mu'}\rangle\langle\nu'\mu'|F|\mu\nu\rangle  +
$$
\begin{equation} \label{14}
\left.\langle \mu|{\bf
S}|\mu'\rangle\langle\nu'\delta\psi_{\mu'}|F|\mu\nu\rangle+\langle \mu|{\bf
S}|\mu'\rangle\langle\nu'\mu'|F|\delta\psi_{\mu}\nu\rangle\right).
\end{equation}
Note, that in the absence of the core polarization the first term
only in Eq. (\ref{14}) contributes into $\delta {\bf S}_0$.  And
this is the only term that produces the Schiff moment of the
nucleus with an odd neutron, like $^{199}$Hg. The residual
interaction $F$ has the form
\begin{equation}\label{15}
F = C\left(f(r) + f'(\mbox{\boldmath $\tau$}_1\cdot
\mbox{\boldmath $\tau$}_2) + g_s(\mbox{\boldmath $\sigma$}_1\cdot
\mbox{\boldmath $\sigma$}_2) + g'_s (\mbox{\boldmath $\sigma$}_1\cdot
\mbox{\boldmath $\sigma$}_2) (\mbox{\boldmath $\tau$}_1\cdot  \mbox{\boldmath
$\tau$}_2)\right)\delta({\bf r}_1-{\bf r}_2),
\end{equation}
where $C=$300 MeV$fm^3$, and $f(r)=f_{ex} +
(f_{in}-f_{ex})\frac{\rho(r)}{\rho(0)}$. Note that the angular
dependence of the operators $\bf S$ and $\delta {\bf S}$ is
completely different. While $S_\mu \sim Y_{1\mu}({\bf n})$, the
induced part $\delta S_\mu$ is a superposition of spin dependent
operators $\sigma_\mu$ and $\{\sigma\otimes Y_2({\bf
n})\}_{1\mu}$. For this reason, different parts of the
interaction (\ref{15}) contribute into renormalization of $\bf S$
and $\delta {\bf S}$.

For a spherical nucleus we can separate the
angular variables and solve the obtained equations in coordinate
space. The equations are
\begin{equation}\label{16}
S^a(r) = S_0^a(r) +\int_0^\infty A^{ab}(r,r')S^b(r')\,dr' ,
\end{equation}
where $a=p,n$ and $S_0^p(r)$  is the radial part of the Schiff
moment operator Eq. (\ref{1}) multiplied by $r$. The
particle-hole propagator $A(r,r')$ was calculated by means of the
Green functions of radial Schr\"odinger equation.
$$
A(r,r') = \frac{C}{3} \cdot Tr_2\{ (f(r) + f'(\mbox{\boldmath
$\tau$}_1\cdot \mbox{\boldmath $\tau$}_2))\sum_{\kappa jl} n_\kappa |\langle
jl||Y_1||\kappa\rangle|^2 rR_\kappa(r)\times
$$
 \begin{equation}\label{17}
r'R_\kappa(r')(G_{jl}(r,r'|\epsilon_\kappa+\omega)+G_{jl}(r,r'|\epsilon_\kappa-
\omega)\}.
\end{equation}
Similar equations can be written for $\delta S(r)$ - the local part of the
induced moment. They differ from Eq. (\ref{17}) in type of tensor operators and
in residual interaction.

 \subsection{Separation of the spurious component}
The integral equation (\ref{16}) must have zero eigenmode related to the center
of mass motion.  However, in our case the mean field Eq. (\ref{6}) is not
consistent with the interaction Eq. (\ref{15}) and, in general, we do not have
zero energy for the center of mass motion. The situation can be improved using
some freedom in the value of the interaction constant $f_{in}$ in Eq.
(\ref{15}).  We can fix the value by the condition $\omega_0 = 0$ for the lowest
isoscalar dipole mode. Since this procedure is numerical the condition
$\omega_0 = 0$ cannot be fulfilled exactly but with finite accuracy. For
$^{199}$Hg we found that the set $f_{in}=0.3935$, $f_{ex}=-2.6$, and $f'=1.07$
gives for $\omega_0$ the value $\omega_0=0.1$ KeV which is really small compared
to the energy of dipole transitions.

The finite accuracy in determination of the spurious mode brings
another problem for solutions of the Eq. (\ref{16}). The bare
operator $S_0(r)$ becomes non orthogonal to the spurious mode
transition density. This results in admixture of the spurious
component to all solutions of the Eq. (\ref{16}). The spurious component
should be subtracted since it can change the solution
considerably. The subtraction can be performed using analytical
properties of the solution as a function of $\omega$. The
solution of the inhomogeneous linear integral equation (\ref{16})
as a function of $\omega$ has the first order poles at $\omega =
\pm E_{ex}$, where $E_{ex}$ are the excitation energies of the
RPA modes. Thus, at small $\omega$ the solution  can be presented
as
\begin{equation}\label{18}
S(r|\omega) = \frac{a(r)}{\omega^2-\omega_0^2} +
b(r|\omega=0)+O(\frac{\omega^2}{E_{ex}^2}).
\end{equation}
The first term in Eq. (\ref{18}) is the spurious component
contribution. Let us define the following set of integrals in a
complex $\omega$-plane along a circle with the radius satisfying
the conditions  $\omega_0\ll |\omega| \ll E_{ex}$.
\begin{equation}  \label{19}
J_n(r) =  \oint  \omega^n  S(r|\omega) \frac{d\omega}{2\pi\imath}.
\end{equation}
Using Eq. (\ref{18}) we can perform the integration analytically As a result we
obtain
$$
 b(r|\omega=0)  =   J_{-1}(r) ,
$$
\begin{equation}  \label{19'}
\omega_0^2 = J_3(r)/J_1(r).
\end{equation}
The numerical integration in Eq. (\ref{19}) has been performed using
32-points Gauss formula. The integration radius was $|\omega|=0.1$ MeV.  The
obtained field $b(r|\omega=0)$ was practically insensitive to the integration
radius. The  value of $\omega_0^2$  was independent on $r$ in first 5 or 6
digits. The loss of accuracy was noticeable only at $|\omega| \sim \omega_0$.

Fig.1 shows the result for renormalized $S(r)=b(r|\omega=0)$ with subtracted
spurious component. It is shown by full line. The dashed line
shows the unrenormalized component $S_0(r)$.  The effects of the
core polarization are not large. We found that for $^{209}$Bi they change the
valence proton contribution by $\sim 15\%$ which is in fair
agreement with the estimates of Ref. \cite{hb00}.  The proton
component does not contribute to the Schiff moment of $^{199}$Hg
since the valence nucleon is a neutron. However, due to
neutron-proton residual strong interaction some neutron component
is produced by the core polarization. This component is shown in
Fig.2.
\begin{figure}
\includegraphics[width=0.47\textwidth ]{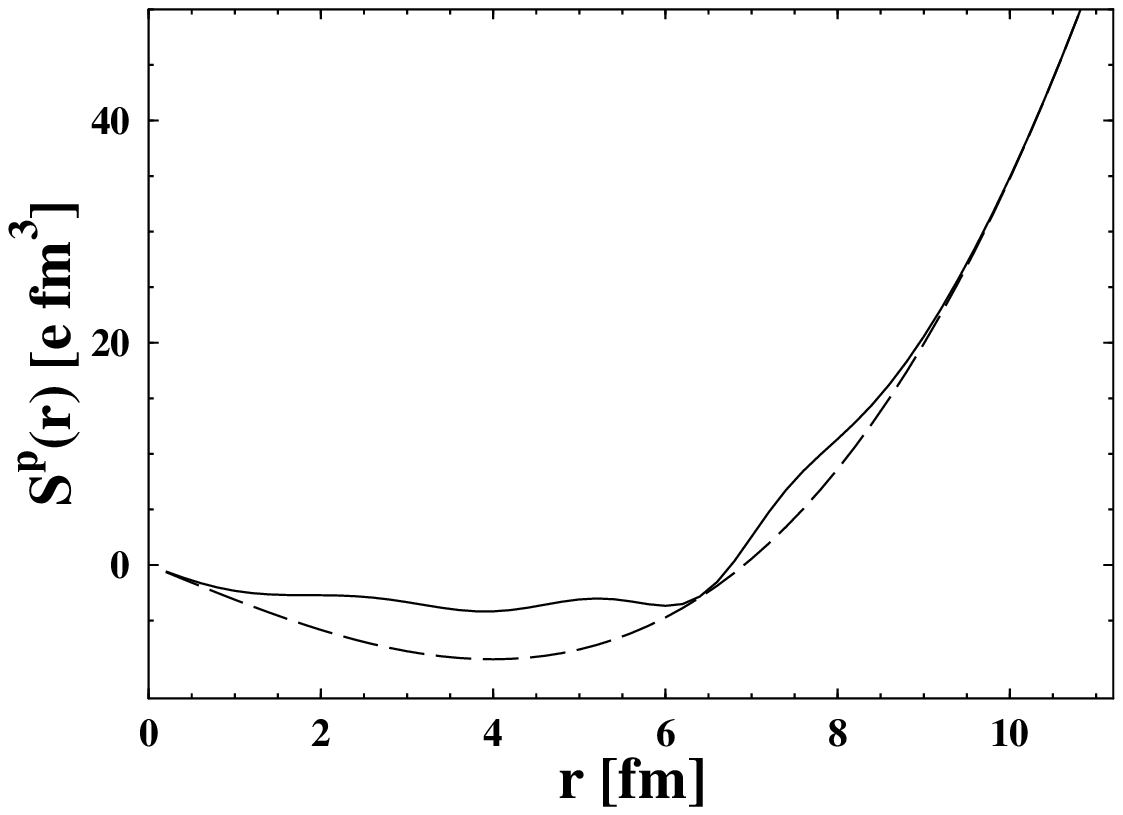}
\hfill
\includegraphics[width=0.48\textwidth ]{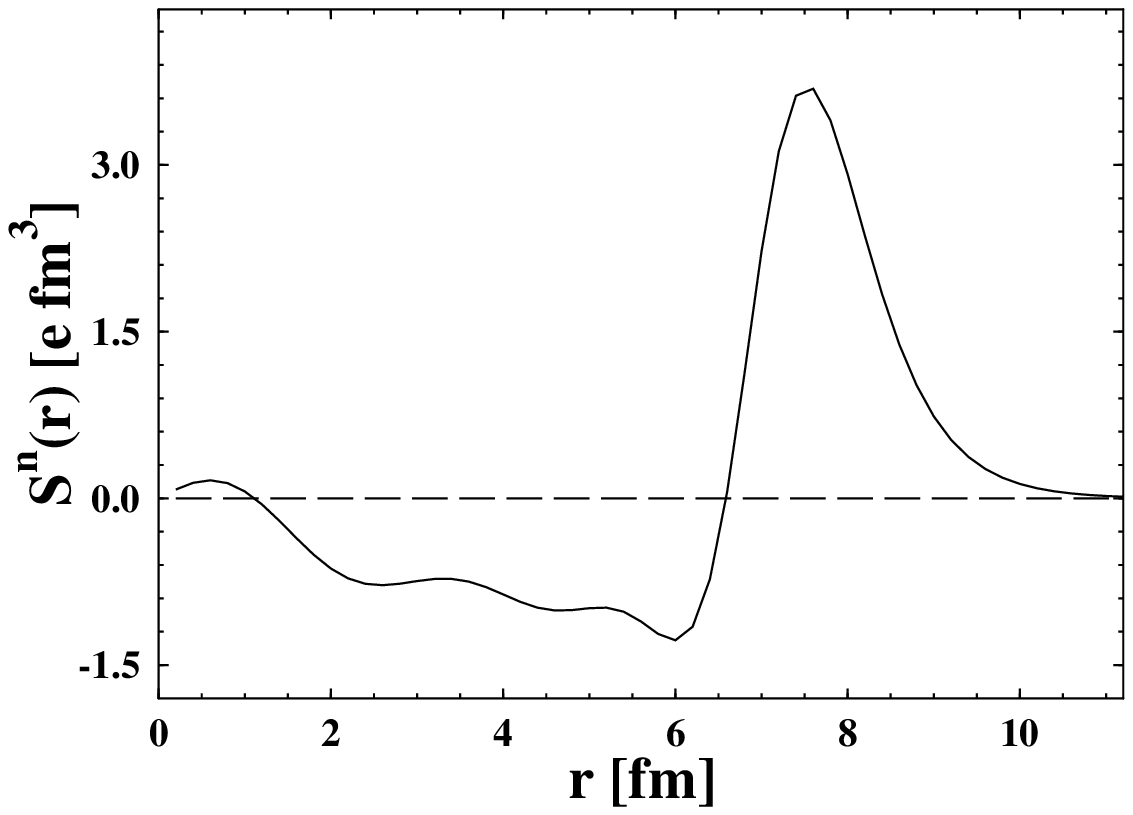}
\\
\parbox[t]{0.47\textwidth}{\caption{The proton component of the Schiff moment.
Full line is the renormalized operator after subtraction of the spurious
component. Dashed line is the bare operator Eq. (1).}}
\hfill
\parbox[t]{0.48\textwidth}{\caption{The neutron component of the
renormalized Schiff moment. }}
\end{figure}
Qualitatively, the behavior of the neutron component inside  a nucleus
resembles the behavior of the bare operator in Fig.1. However, the magnitude of
the neutron component is smaller and the behavior outside of the nucleus is
completely different.
\section{Results for $^{199}$Hg }
In Table 1 we show the contributions of the neutron component discussed above to
the Schiff moment of  $^{199}$Hg .
\begin{table}[h]
\caption{Contributions of the neutron component to the Schiff moment of
$^{199}$Hg in [e fm$^3$].}
\begin{center}
\begin{tabular}{||c|c|c|c||}
\hhline{|t:=:=:=:=:t|}
& $gg_0 $ &  $gg_1 $ &  $gg_2$  \\
&&& \\
\hhline{||----||}
direct & 0.0038 & -0.024 & -0.0076 \\
 \hhline{||----||}
exchange &0.0032 & -0.002 & -0.0015 \\
 \hhline{||----||}
total & 0.0070 & -0.026 & -0.0091 \\
 \hhline{|b:=:=:=:=:b|}
\end{tabular}
\end{center}
\end{table}
The three columns correspond to three isospin channels. The first
row is the contribution from $\delta R_{dir}$ produced by $\delta
U_{dir}$ given by Eq. (\ref{7}). The second row is the contribution from
$\delta R_{exch}$  produced by  $\delta U_{exch}$ given by Eq. (\ref{9}),
and the third row is the sum of them.

In the direct contributions the isospin channel $T=1$ dominates, as it was
mentioned above. The exchange contributions are small, in general. For the
isospin channels $T=0,2$ they are comparable to the direct contributions just
because the llatter's are suppressed by the factor $(N-Z)/A$.

The contributions from $\delta {\bf S}$ are more significant. In Table 2 we list
the contributions from the direct and the exchange parts of the weak interaction
$W$.
 \begin{table}[h]
\caption{Induced contributions to the Schiff moment of $^{199}$Hg in [e
fm$^3$].}
\begin{center}
\begin{tabular}{||c|c|c|c||}
\hhline{|t:=:=:=:=:t|}
& $gg_0 $ &  $gg_1 $ &  $gg_2$  \\
&&& \\
\hhline{||----||}
direct $\delta {\bf S}_0$& -0.0302 & -0.0631 & 0.0604 \\
 \hhline{||----||}
exchange $\delta {\bf S}_0$& -0.0007& -0.0012 & -0.0007 \\
 \hhline{||----||}
direct $\delta {\bf S}$ & -0.0086 & -0.0285 & 0.0172 \\
 \hhline{||----||}
 exchange $\delta {\bf S}$& -0.0002 & -0.0008 & -0.0003\\
 \hhline{||----||}
non-local  $\delta {\bf S}_{NL}$ & 0.0014 &-0.00004 &0.0013 \\
 \hhline{|b:=:=:=:=:b|}
\end{tabular}
\end{center}
\end{table}
In the first two rows the unrenormalized contributions $\delta
{\bf S}_0$ are shown. Again, the exchange contributions are
considerably smaller than the direct ones. The contributions from
the channels $T=1$, and $T=2$ are comparable here although $T=1$
contribution is still larger. Next two rows show the renormalized
contributions $\delta {\bf S}$. The effect of the core
polarization is significant here. The induced moment   $\delta
{\bf S}$ includes spin dependent operators, therefore, the
spin-spin part of the residual interaction Eq. (\ref{15}) is
responsible for their renormalization. The spin-spin interaction
is repulsive with the constants $g_s=0.63$, and $g'_s=1.01$. The
repulsion results in decrease of absolute values of the
renormalized contributions. Finally, in the last row we show the
contribution of the non-local term  $\delta {\bf S}_{NL}$ Eq.
(\ref{14}). It has the exchange origin as well and its
contribution is really insignificant. All the contributions can
be summarized as follows
 \begin{equation}\label{20}
 S =    -0.0004 gg_0  - 0.055 gg_1 + 0.009 gg_2\; [e{\rm fm}^3].
\end{equation}
The obtained value for the Schiff moment in Eq. (\ref{20}) cannot be compared
directly with the previous calculations with the contact interaction Eq.
(\ref{5}). The reason is in different definition of the dimensionless constants
$gg_i$ in Eq. (\ref{4})  and $\eta_{ab}$ in Eq. (\ref{5}). To perform the
comparison we redefine the constants $g_i $
 \begin{equation}\label{21}
g_i= \frac{Gm_\pi^2}{\sqrt{2}} \tilde{g}_i.
\end{equation}
With this factor the integration over space of the Yukawa function gives $1$,
exactly as the integration of a $\delta (\bf r)$.  Introducing this factor we
obtain
\begin{equation}\label{22}
S = (-0.01 g\tilde{g}_0 -0.86 g\tilde{g}_1 + 0.14 g \tilde{g}_2
)\times 10^{-8} \; [e\,{\rm fm}^3].
\end{equation}
This value should be compared with $S= -1.4\times 10^{-8} \eta_{np}$ from Ref.
\cite{fks86}, and  $S\approx -1.6\times 10^{-8} \eta_{np}$  from Ref.
\cite{fg02}.  Remembering that $\eta_{np} \sim g(\tilde{g}_0 +\tilde{g}_1
- 2\tilde{g}_2) $ we conclude that the difference between our result and
previous calculations is significant for $T=0$, and $T=2$ channels. Our values
are smaller in absolute value. In order to trace
the origin of this difference we repeated our calculations using the contact
interaction and omitting the core polarization. The contact interaction was
obtained by replacing the Yukawa function in Eq. (\ref{4}) by the
delta-function.  The result is
\begin{equation}\label{23}
S=-0.96\times 10^{-8}g(\tilde{g}_0+\tilde{g}_1-2 \tilde{g}_2)  \; [e\,{\rm
fm}^3].
\end{equation}
If we omit completely the effect of the core polarization in
calculations with the finite range interaction Eq. (\ref{4}) then
the only nonzero contribution in the first row of  the Table 2
becomes
$$
S= -0.086\, g(g_0+g_1 -2 g_2)\; [e\,{\rm fm}^3],
$$
that corresponds to
\begin{equation}\label{24}
 S=-1.35\times 10^{-8}g(\tilde{g}_0+\tilde{g}_1-2 \tilde{g}_2)  \; [e\,{\rm
fm}^3].
\end{equation}
Comparing  Eq. (\ref{23}) and Eq. (\ref{24}) we conclude that the
effect of finite weak interaction range is not very significant.
The main effect bringing the value of the Schiff moment from
that in Eq. (\ref{24})  to the value in Eq. (\ref{22}) comes from
the core polarization.

The last remark concerns the pairing effects.    The nucleus $^{199}$Hg has 7
neutron holes in the unfilled shell. One of them is fixed in $p_{1/2}$ state and
6 others should be distributed among the other states in the shell. From mass
differences we found $\Delta_n = 0.69$ MeV. This value of the pairing
gap is typical for developed pairing.  For dipole transitions the transition
energy is large compared to  $\Delta$ and the pairing effects are small. They
were omitted in Eq. (\ref{16}).
 For the induced moment the situation is different. There, the transitions with
$\Delta J=0,1$ and $\Delta L=0,2$  are responsible for the core polarization.
Such transitions exist inside the last unfilled neutron shell and, due to
Pauli blocking, they are sensitive to the details of the shell occupation.
For a T-odd operator the effects of pairing in the core polarization can be
considered  in the way used in Ref. \cite{dt97}.  We found, that the main effect
of pairing is in fixing the occupation numbers in the upper unfilled
neutron shell. As soon as we keep the occupation numbers fixed the results for
$\delta {\bf S}$ are changed within few percents when we put $\Delta_n = 0$.

In summary, we calculated the Schiff moment of $^{199}$Hg nucleus
using finite range weak interaction and considering the core
polarization effects. The effects of the finite interaction range are not
very significant for the Mercury nucleus. They do not change the
order of magnitude of the Schiff moment calculated with the
contact interaction.  The effects of the core polarization are
two-fold. First, they renormalize the bare operator of the Schiff
moment producing a small neutron component due to n-p residual
interaction. Second, they produce an induced Schiff moment due to
P- and T- violation component in the intermediate single particle
states. The induced moment is proportional to both the strong
residual and the weak interactions. The effects of the core
polarization for the Mercury nucleus are large and they have to
be accounted in calculations of P- and T-violating effects in the
Mercury nucleus.
\section*{Acknowledgments}
The authors appreciate discussions with I.B. Kriplovich during this work.


\begin{thebibliography}{99}
\bibitem{rgf01} M.V. Romalis, W.C. Griffith, and E.N. Fortson, Phys. Rev. Lett.
 {\bf 86}, 2505, (2001).
\bibitem{pr50} E.M. Purcell and N.F. Ramsey, Phys. Rev.  {\bf 78}, 807,
(1950).
\bibitem{sch63} L.I. Schiff, Phys. Rev.  {\bf 132}, 2194, (1963).
\bibitem{san67} P.G.H. Sandars, Phys. Rev. Lett.  {\bf 19}, 1396, (1967).
\bibitem{fks84} V.V. Flambaum, I.B. Khriplovich, and O.P. Sushkov, Zh. Eksp.
Teor. Fiz.  {\bf 87}, 1521, (1984) [Sov. Phys. JETP  {\bf 60}, 873, (1984)].
\bibitem{fg02} V.V. Flambaum and J.S.M. Ginges, Phys. Rev. A {\bf 65},
032113, (2002).
\bibitem{efh00}  J. Engel, J.L. Friar, and A.C. Hayes, Phys. Rev. C {\bf 61},
 035502, (2000).
\bibitem{hb00} I. Hamamoto and B.A. Brown, Phys. Rev. C {\bf 62},
024318, (2000).
\bibitem{hh83}W.C. Haxton and E.M. Henley, Phys. Rev. Lett. {\bf 51},
1937, (1983) .
\bibitem{h88} P. Herczeg, Hyperfine Interactions {\bf 43}, 77, (1988).
\bibitem{khk00} I.B. Khriplovich and R.V. Korkin, Nucl. Phys. A {\bf 665},
365, (2000).
\bibitem{kh91} I.B. Khriplovich, {\it Parity Nonconcervation in Atomic
Phenomena}, (Gordon and Breach, Philadelphia, 1991)
\bibitem{bs74} B.L. Birbrair and V.A. Sadovnikova, Yad. Fiz.
{\bf 188}, 1851, (1974).
\bibitem{m67} A.B. Migdal, {\it Theory of Finite Fermi System},
(Wiley, New York, 1967).
\bibitem{dt97} V.F. Dmitriev and V.B. Telitsin, Nucl. Phys. A {\bf 613}, 237,
(1997), V.F. Dmitriev and V.B. Telitsin, Nucl. Phys. A {\bf 674}, 168, (2000).
\bibitem{fks86} V.V. Flambaum, I.B. Khriplovich, and O.P. Sushkov, Nucl. Phys. A
{\bf 449}, 750 (1986).
\end{thebibliography}
\end{document}